\documentclass[conference]{IEEEtran}

\ifCLASSINFOpdf
\else
\fi
\usepackage{nomencl}
\usepackage{url}
\usepackage{ifpdf}
\usepackage{cite}
\usepackage{wrapfig}
\usepackage{graphicx}
\usepackage{dirtytalk}
\usepackage{mathtools}
\usepackage{cite}
\usepackage{amsmath,amssymb,amsfonts}
\DeclareMathOperator*{\argmax}{arg\,max}
\usepackage{textcomp}
\usepackage{bigints}
\usepackage{xcolor}
\usepackage{tikz}
\usepackage{float}
\graphicspath{ {./Figures/} }
\usepackage{bbm}
\usetikzlibrary{shapes.geometric, arrows}
\usepackage{algorithmic}
\usepackage[ruled,vlined]{algorithm2e}
\usepackage{array}
\ifCLASSOPTIONcompsoc
 \usepackage[caption=false,font=normalsize,labelfont=sf,textfont=sf]{subfig}
\else
\usepackage[caption=false,font=footnotesize]{subfig}
\fi
\usepackage{dblfloatfix}
\ifCLASSOPTIONcaptionsoff
 \usepackage[nomarkers]{endfloat}
\let\MYoriglatexcaption\caption
\renewcommand{\caption}[2][\relax]{\MYoriglatexcaption[#2]{#2}}
\fi
\let\MYorigsubfloat\subfloat
\renewcommand{\subfloat}[2][\relax]{\MYorigsubfloat[]{#2}}
\usepackage{url}
\usepackage[]{fancyhdr} %
\newcommand{\changefont}{\fontsize{9}{9}\selectfont}
\IEEEoverridecommandlockouts
\makenomenclature
\begin{document}

%
\title{On Statistical Modeling of Load in Systems with High Capacity Distributed Energy Resources}

\author{\IEEEauthorblockN{Aaqib Peerzada}
\IEEEauthorblockA{ECE Department\\ Texas A\&M University\\College Station, TX, USA\\
peerzada@tamu.edu}
\and
\IEEEauthorblockN{Miroslav Begovic\\}
\IEEEauthorblockA{ECE Department\\ Texas A\&M University\\College Station, TX, USA\\
begovic@tamu.edu} 
\and
\IEEEauthorblockN{Wesam Rohouma\\}
\IEEEauthorblockA{ECE Department\\ College of North Atlantic\\ Doha, Qatar\\
wesam.rohouma@cna-qatar.edu.qa}
\and
\IEEEauthorblockN{Robert S. Balog\\}
\IEEEauthorblockA{ECE Department \\ Texas A\&M University at Qatar\\ Doha, Qatar\\
robert.balog@qatar.tamu.edu}
}


%





\maketitle
\thispagestyle{fancy}
\pagestyle{fancy}


\begin{abstract}
The emergence of distributed energy resources has led to new challenges in the operation and planning of power networks. Of particular significance is the introduction of a new layer of complexity that manifests in the form of new uncertainties that could severely limit the resiliency and reliability of a modern power system. For example, the increasing adoption of unconventional loads such as plug-in electric vehicles can result in uncertain consumer demand patterns, which are often characterized by random undesirable peaks in energy consumption. In the first half of 2021, the  electric vehicle sales increased by nearly 160\%, thus accounting for roughly 26\% of new sales in the global automotive market. This paper investigates the applicability of generalized mixture models for the statistical representation of aggregated load in systems enhanced with high capacity distributed energy resources such as plug-in electric vehicles.
\end{abstract}



%
\printnomenclature
\IEEEpeerreviewmaketitle

\nomenclature{\(N_{EV}(t)\)}{Number of EV arrivals in $[0,t]$}
\nomenclature{\(\Omega\)}{Sample Space on which the Counting Process is defined}
\nomenclature{\(S_{k}\)}{$k^{th}$ Arrival Epoch of Homogeneous Poisson Process}
\nomenclature{\(\lambda(t)\)}{Time Varying Intensity Function}
\nomenclature{\(\wedge (t)\)}{Mean Value Function}
\nomenclature{\(X_{k}\)}{$k^{th}$ Inter-arrival time}
\nomenclature{\(F(.)\)}{CDF Function}
\nomenclature{\(\lambda^{+}\)}{Constant Rate Intensity Function}
\nomenclature{\(S_{k}^{NHPP}\)}{$k^{th}$ Arrival Epoch of Nonhomogeneous Poisson Process}
\nomenclature{\(SoC_{Arrival}\)}{State of Charge of electric vehicle battery at Arrival}
\nomenclature{\(E_{req}\)}{Energy Required to charge electric vehicle battery}
\nomenclature{\(T_{ch}\)}{Time required to charge electric vehicle battery}
\nomenclature{\(P\)}{L-2 Charging Rate of electric vehicle}
\nomenclature{\(f_{Y}(.)\)}{Probability Density Function}
\nomenclature{\(\pi_j\)}{Weight assigned to $j^{th}$ component density function}
\nomenclature{\(\Psi\)}{Vector of Model Parameters}
\nomenclature{\(\bold Z\)}{Vector of Hidden Variables}
\nomenclature{\(\bold Y\)}{Vector of Load Measurements}
\nomenclature{\(\beta\)}{Shape Parameter of Generalized Gaussian Distribution}
\nomenclature{\(s\)}{Scale Parameter of Generalized Gaussian Distribution}
\nomenclature{\(\mu\)}{Location Parameter of Generalized Gaussian Distribution}
\nomenclature{\(\epsilon\)}{Tolerance level}
\nomenclature{\(\bold S\)}{Vector of arrival epochs}
\nomenclature{\(d\)}{Daily Driven Miles}
\nomenclature{\(E_{cons}\)}{Electricity consumption in kWh/100 miles}
\nomenclature{\(C_b\)}{EV Battery Capacity in kWh}

\section{Introduction}

At the end of the year 2020, there were about 10 million electric vehicles (EVs) on the roads globally \cite{IEA2020GlobalIEA}. Despite the uncertainty and the disruption caused by COVID-19 in the global supply chain, the EV registrations increased by 41\% in 2020, with Europe superseding the People's Republic of China as the world's largest  EV market \cite{IEA2020GlobalIEA}. The robust nature of the EV market is primarily due to the supportive regulatory frameworks that many countries have adopted, such as a significant reduction in carbon dioxide emissions and zero-emission vehicles mandates. The improvements in battery technology and a continued decline in battery costs are other contributing factors that explain EV sales' resiliency. In the U.S, for instance, the federal government has set out a \$ 174 million commitment to support the adoption of EVs, with  President Biden signing an executive order in August 2021 to have EVs make up for nearly 50\% of all the automotive sales by 2030 \cite{Room2021FactSheet}. \par
While the use of EVs may result in an overall benefit both in terms of direct and Well-to-Wheel emissions, when seen from a power system perspective, such unconventional loads may present significant security and reliability challenges to the normal operation of the electric grid. The reliability concerns include thermal overloading of power transformers and violation of transmission line capacity limits, while the security challenges include increased electric demand with a pronounced  \say{peaky} behavior, increased power loss, and injection of harmonics into the grid \cite{Elnozahy2014ANetworks}. Also, the uncertain nature of the connection of the EVs to the electric grid makes load forecasting more challenging due to the introduction of new random consumer demand patterns. Furthermore, the uncoordinated charging, particularly the fast three-phase charging, could increase the instances of thermal violations in lines, cables, and transformers, and the impact can exacerbate if it coincides with the peak energy consumption \cite{Clement-Nyns2010TheGrid},\cite{Muratori2018ImpactDemand}\cite{Hilshey2013EstimatingAging}. \par 
The assessment of some of the damaging effects of EV charging heavily depends on the mathematical model used to model EVs arrival at a charging facility. The EV charging demand is a stochastic process and the expected value of the random variable representing EV demand can be calculated based on the expected number of the EV vehicles queuing up at a charging station to receive charging service in a given time period. This necessitates modeling the EV charging as a stochastic counting process. The literature on EV modeling as a counting process is scarce and limited to models based on queuing theory.  An important study on stochastic EV modeling is presented in \cite{Alizadeh2014AVehicles}. In the study the authors utilize a stochastic model based on queuing theory to forecast the EV demand profile using real-time sub-metering data. The EV arrival process is treated as a point process with random arrival times. The study however does not provide any information on the simulation of the counting process to estimate the total EV charging demand which is important in situations when the real-time sub-metering data is not available.  Another important study \cite{Hafez2018QueuingOperation} uses a  nonhomogeneous Poisson process for the arrival rate of EVs and the intensity function is chosen on the real-world data that  based  either on customer convenience or EV charging price. The vehicle arrival data is directly used to generate the expected EV demand based on the random sampling of the number of vehicles that are being charged simultaneously. A drawback of the method presented in \cite{Hafez2018QueuingOperation} is the lack of information on the arrival times of the EVs. The information about the EV arrival times is critical to account for the temporal dependence of the EV demand. Queuing theory to model the EV demand is also used in \cite{Vlachogiannis2009ProbabilisticVehicles},\cite{Garcia-Valle2009LetterStudies},\cite{Bae2012SpatialDemand}. However, these studies utilize a homogeneous Poisson process with constant arrival rate to model EV demand . The constant arrival rate assumption does not hold as is clear from the National Household Travel Survey (NHTS) data \cite{Bricka2014NationalSurvey}.    \par The power system load is one of the most noticeable operational parameters with a strong temporal dependence. The literature on statistical modeling of load has established that load patterns are highly variable when measured at different buses in a power network. In particular, the study in  \cite{Singh2010StatisticalModel} demonstrates the multi-modal characteristics of the load and proposes the use of a Gaussian mixture model (GMM) for statistical modeling of the load.  A GMM is a powerful computational tool that can be used to fit a probability density function with more than one mode. The study in  \cite{Valverde2012ProbabilisticModels} uses a GMM  to approximate non-Gaussian density functions such as correlated wind power output and aggregated load in the presence of non-Gaussian correlated random input variables.  In \cite{Stephen2014EnhancedCustomers} the authors evaluate the performance of the GMM and  Mixture of Factor Analyzers (MFA) method in modeling residential loads, and the results are compared with the existing load models. The study concludes that both GMM and MFA offer superior performance characteristics compared to the existing British load model. The study in \cite{Quiros-Tortos2018StatisticalApplications} uses probability density functions based on GMM to statistically quantify key charging metrics of EVs. The study uses real data from 221 EVs, part of the largest trial in Europe and the UK. A key constraint in the application of GMM is that the area under the curve of each component density must equal unity over the entire sample space. The study in \cite{Cui2018StatisticalModel} proposes a slightly different version of the GMM to fit the density functions of wind power ramping. The proposed model differs from the conventional GMM model in the sense that the integral of each component density over the entire sample space is not required to be unity. Also, the associated weights of each component density can be negative as opposed to a conventional GMM, where each component density weight must be nonnegative.  However, all the studies focused on statistical load modeling based on GMM ignore the impact of distributed energy resources (DERs), for example, distributed generation and the use of unconventional loads such as plug-in EVs. Due to the uncertainty associated with the use of DERs, particularly roof-top solar generation and EVs, the overall effect is an increase in the \say{peakiness} of the DER-impacted load profile as compared to the traditional load. For this reason we propose the use of a generalized version of the GMM with additional parameters that can be used to control the shape of the distribution function.  \par

Our contributions in this paper are twofold. First, we propose a stochastic counting process based on the nonhomogeneous Poisson process (NHPP) to model the EV demand and present an algorithm to simulate the arrival times of the EVs. The simulation algorithm is based on a version of the acceptance-rejection called \say{thinning}. The charging times of the EVs are estimated from the daily driven miles taken from the National Household Travel Survey (NHTS) data \cite{Bricka2014NationalSurvey} and the battery state of charge at the arrival. Secondly, we propose a generalized version of the GMM for statistical modeling of the load taking into account the impact of EV charging.  The parameters of the proposed mixture model are estimated using the Expectation-Maximization algorithm (E-M) \cite{Sammaknejad2019AIdentification}. We present the update equations of the model parameters and fit the model output to the measured load data considering high-level (L-2) EV charging.

\section{Mathematical Modeling of Electric Vehicles}
In this paper we model the EV connection to the grid as a stochastic counting process $\{N_{EV}(t);t\geq0\}$ defined on a sample space $\Omega$. The function $N_{EV}(t)$ is the realization of the number of events in the interval $[0,t]$. The counting process $\{N_{EV}(t); t\geq0\}$ for any arrival process has the property that $N_{EV}(\tau)\geq N_{EV}(t)$ for all $\tau \geq t$. This means that $N_{EV}(\tau)-N_{EV}(t)$ is a nonnegative random variable. Hence by definition $\{N_{EV}(t); t>0\}$ is integer-valued, non-decreasing and right continuous. The $k^{th}$ arrival epoch $S_k$ is related to the counting random variable $N_{EV}(t)$ 
\begin{equation}
    \{S_k\leq t\}=\{N_{EV}(t)\geq k\}
\end{equation}
This can be verified by observing that the event $\{S_k\leq t\}$ refers to the $k^{th}$ arrival by time $t$. This further implies that the number of events by time $N_{EV}(t)$ must be at least $k$. Conversely it is also true that 
\begin{equation}
    \{S_k>t\}=\{N_{EV}(t)< k\}
\end{equation}

\subsection{EV Connection as an Arrival Process}
The arrivals of EVs at a charging station can be considered a point process with a series of random arrival times. Thus we consider a non-homogeneous Poisson process with an intensity function $\lambda(t)$. The study in \cite{Alizadeh2014AVehicles} has performed a null hypothesis test on the vehicle travel patters based on the National Household Travel Survey (NHTS) data and it is shown that the EV arrivals for charging can be modeled as a constant-rate Poisson process in short-intervals of time (30 minutes).  The counting process $\{N_{EV}(t); t>0\}$  is a non-homogeneous Poisson process with time varying arrival rate of $\lambda(t)$ and has the independent increment property. In addition, $\forall t\geq0$ and $\delta>0$, $\{N_{EV}(t); t>0\}$ satisfies \cite{Papoulis1967Processes}
\begin{equation}
\begin{aligned}
    &\Pr\{\Tilde{N}_{EV}(t,t+\delta)=0\}=1-\lambda(t)\delta+o(\delta^2)\\
    &\Pr\{\Tilde{N}_{EV}(t,t+\delta)=1\}=\lambda(t)\delta+o(\delta^2)\\
    &\Pr\{\Tilde{N}_{EV}(t,t+\delta)\geq2\}=o(\delta^2) \label{eq:2.1}
\end{aligned}
\end{equation}
In (\ref{eq:2.1}), $\Tilde{N}_{EV}(t,t+\delta)=N_{EV}(t+\delta)-N_{EV}(t)$. The non-homogeneous Poisson process (NHPP) defined in (\ref{eq:2.1}) does not have the stationary increment property. $\lambda(t)$ is also called the rate function of the NHPP.  The NHPP (\ref{eq:2.1}) is  characterized by the mean value function $\wedge(t)\equiv \mathbb{E}[N_{EV}(t)]$. The mean value function in terms of the intensity function $\lambda(t)$ is
\begin{equation}
    \wedge(t)=\int_{0}^{t}\lambda(y)dy <\infty
\end{equation}
For a NHPP, the probability of having $k$ arrivals in the interval $[0,t]$ is given by \cite{Papoulis1967Processes}
\begin{multline}
    \Pr\{N_{EV}(t)-N_{EV}(0)=k\}=\\ \frac{\left[\wedge(t)-\wedge(0)\right]^k}{k!}\exp(-\left[\wedge(t)-\wedge(0)\right]) \label{eq:2.2}
\end{multline}

\subsection{Generating Non-homogeneous Poisson Process}
The inter-arrival times of a NHPP are not independent and do not have an exponential distribution unlike a homogeneous Poisson process. Specifically, the cdf of the $k^{th}$ inter-arrival time $X_k=S_{k+1}-S_{k}$ conditional on the arrival epochs $\{S_j; j=1,2,...,k\}$ is  
\begin{equation}
    F_{s_k}(x)=1-\exp(-\wedge (s_{k}+x)+\wedge (s_{k})) \label{eq:2.3}
\end{equation}
Equation (\ref{eq:2.3}) can be derived using the independent increment property of a NHPP (i.e.) we consider the random variables $\{N_{EV}(I_{n}); 1\leq n\leq k\}$ to be independent where $\{I_n\}_{1\leq n\leq k}$ are piece wise disjoint intervals. \par
A NHPP can be generated from a homogeneous Poisson process by considering a constant intensity function $\lambda^{+}$ that dominates the time-varying intensity function $\lambda(t);t\geq0$ of the desired NHPP such that $\lambda^{+}\geq \lambda(t)$ $\forall t \in [0,T]$. A variation of the acceptance-rejection called "thinning" is used to sample from the generated events of a homogeneous Poisson process such that the desired intensity function $\lambda(t)$ is achieved \cite{Pasupathy2011GeneratingProcesses}. The thinning algorithm is based on the following theorem \par
$\textbf{Theorem 1} $ (Lewis and Shedler, 1979 \cite{Lewis1979SIMULATIONTHINNING.}) Consider a non-homogeneous Poisson process with intensity function $\lambda_{v}(t),t\geq0$. Suppose that $S_{1}^{*}, S_{2}^*,...,S_{k}^*$ are random variables representing event times from the non-homogeneous Poisson process with intensity function $\lambda(t)$ and lying in the fixed interval $(0,t]$. Let $\lambda(t)$ be a intensity function such that $0\leq\lambda(t)\leq\lambda_{v}(t) \forall t\in [0,t]$. If the $i^{th}$ events is independently deleted with probability $1-\lambda(t)/\lambda_{v}(t)$, the remaining event times form a nonhomogeneous Poisson process with intensity function $\lambda(t) $ in the interval $(0,t]$. The proof is given in \cite{Chen2016ThinningProcesses} \par
The thinning algorithm used in the paper to simulate a nonhomogeneous Poisson process is implemented as follows. We consider $\lambda(t)$ to be the intensity function of the NHPP over a fixed interval $[0,T]$. \par
\begin{itemize}
    \item Simulate a homogeneous Poisson process (HPP) with constant intensity function $\lambda^+\geq \lambda(t) \forall t\in[0,T]$ by drawing uniform random numbers $\{u_k;k=1,2,...,k^*\} \sim U(0,1)$. Since the inter-arrival times are exponentially distributed in a HPP, the inter-arrival times are obtained by setting $X_k=-\frac{1}{\lambda^{+}}\log u_k$. 
    \item The arrival times of a HPP are obtained by setting $S_k=S_{k-1}-\frac{1}{\lambda^{+}}\log u_k$. The total number of uniform random numbers drawn is $k^*=\max \{k;\sum_{n=1}^{k}S_n<T \}$.
    \item Independently generate uniform random numbers $\{w_j; j=1,2,...,k^*\}\sim U(0,1)$ and calculate the indicator function
    \begin{equation}
  I_{j}=\begin{cases}
    1; w_j\leq \frac{\lambda(S_j)}{\lambda^{+}} \label{eq:2.4} \\
    0; w_j> \frac{\lambda(S_j)}{\lambda^{+}} 
  \end{cases}
\end{equation}
\item Form the set $\mathbf{J}=\{I_j;j=1\}$ and the arrival times of the NHPP are $S_j^{NHPP}=\{S_j;j\in \mathbf{J}\}$
\end{itemize}
\begin{algorithm}
\caption{Acceptance Rejection based Thinning of HPP}
\begin{algorithmic}
\STATE \textbf{Input}:NHPP Intensity function $\lambda(t)$, HPP Constant-rate Intensity function $\lambda^{+}$, Interval length $[0,T])$ 
\STATE Set $\lambda^{+}=\max \{\lambda(t)\}$
\STATE Set the counting process $N_{EV}=a\lambda^{+}T, a>1$
\WHILE{$i\leq N_{EV}$}
\STATE Draw $u\sim U(0,1)$
\STATE Set $S_k=-\frac{1}{\lambda^{+}}\log u_k$
\ENDWHILE
\STATE  Set $\bold S=\bold S (\bold S <N_{EV})$
\STATE Set $k^*=\max \{k;\sum_{n=1}^{k}S_n<T \}$
\FORALL{i=1,..,k}
\STATE Draw $w\sim U(0,1)$
\STATE Calculate acceptance probability, $ r(j)=\frac{\lambda(S_j)}{\lambda^{+}}$
\IF{$w_j \leq r(j)$}
\STATE $\bold I_j=1$
\ELSE
\STATE  $\bold I_{j}=0$
\ENDIF
\ENDFOR
\STATE $\mathbf{J}=\{\bold I_{j}; j=1\}$
\STATE $S_j^{NHPP}=\{S_j;j\in \mathbf{J}\}$
\STATE \textbf{Output}: Arrival Times of NHPP
\end{algorithmic}
\end{algorithm}

\begin{figure}[H]
    \centering
    \hbox{\hspace{-2.0em}}\includegraphics[width=3.5in,height=2.0in]{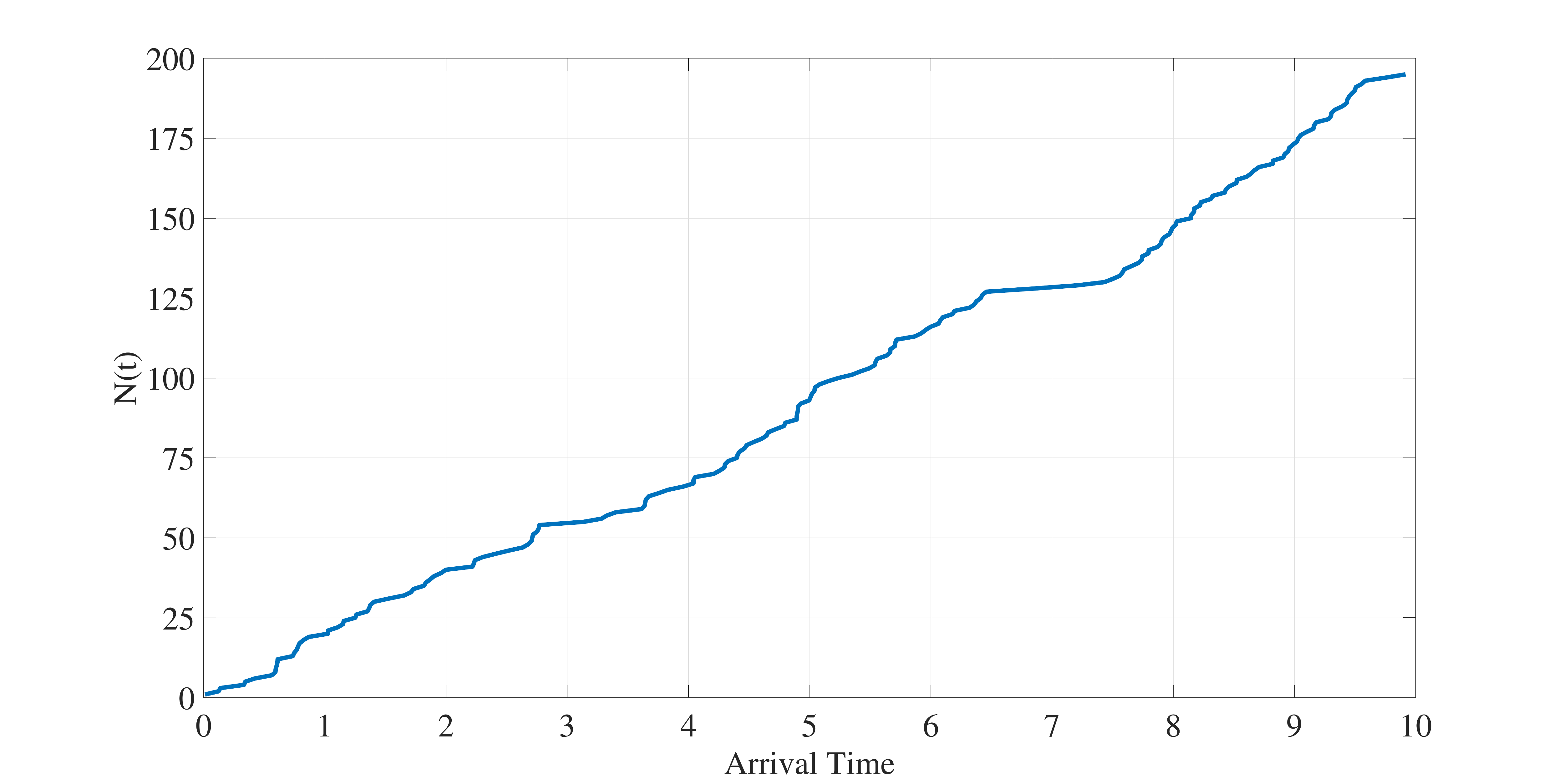}
    \vspace{-2.0em}\caption{Count of NHPP with $\lambda(t)=20+10\sin(0.5\pi t)$}
    \label{fig:1}
\end{figure}

\begin{figure}[H]
    \centering
    \hbox{\hspace{-2.0em}}\includegraphics[width=3.5in,height=2.0in]{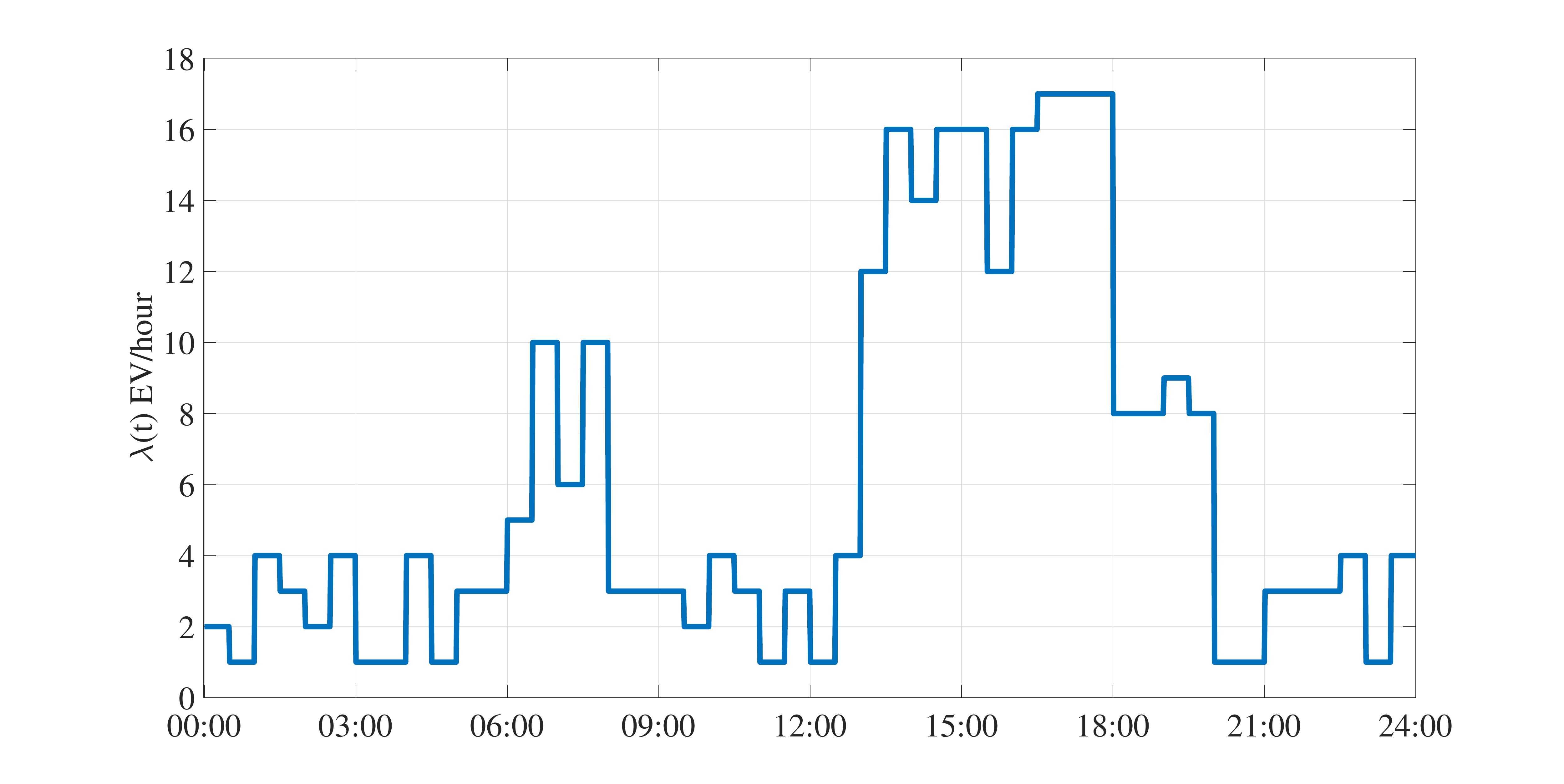}
    \vspace{-2.0em}\caption{Time-varying Arrival Rate}
    \label{fig:3}
\end{figure}

To verify the accuracy of the generation procedure based on thinning, we simulate a NHPP with a continuous time varying intensity function $\lambda(t)=20+10\sin(0.5\pi t)$ over a fixed interval $[0,10]$. The arrival times of the NHPP are sampled from the vector of arrival times of the HPP with intensity function $\lambda^{+}=30$. The simulation is repeated 10000 times and the mean of the simulated counting process is compared with the theoretical mean which is given by the mean value function evaluated at the end points of the fixed interval. A plot of the number of arrivals as a function of arrival times is shown in Fig \ref{fig:1} and results are presented in Table \ref{tab1} 
\vspace{-2.0em}
\begin{table}[H]
\centering
\caption{Theoretical and Simulated Mean Values}
\label{tab1}
\vspace{-1.0em}
\begin{tabular}{|c|c|}
\hline
Theoretical Mean Value& 213.09         \\ \hline
Simulated Mean Value&212.73  \\ \hline  
\end{tabular}
\end{table}
In Fig \ref{fig:1} it is clear that the process trajectory has a sinusoidal pattern consistent with the assumed intensity function. The theoretical mean value can be calculated from the mean value function $\wedge (t)$ evaluated at the end point of the fixed interval and is given by the integral 
\begin{equation}
    \wedge(t)=\int_0^{t}20+10\sin(0.5\pi y)dy
\end{equation}
To estimate the arrival times of  EV arrival process we consider a NHPP with a piecewise constant intensity function for intervals lasting 30 minutes. The piecewise constant intensity function depends on the customer convenience and is based on real world data given in \cite{Hafez2018QueuingOperation}. For a typical day the piecewise constant arrival rate is given in Table \ref{tab2}
\begin{table}[H]
\centering
\caption{Arrival Rate of EV Connection \cite{Hafez2018QueuingOperation}}
\label{tab2}
\vspace{-1.0em}
\begin{tabular}{|c|c|c|}
\hline
Time (Mins)& \% of Vehicles on Road & EVs/hour         \\ \hline
$0\leq t<360$& $\leq 4$& [1,2,3,4]  \\ \hline  
$360\leq t<480$& $>$4 and $\leq 7$ & [5,6,7,8,9,10,11]      \\ \hline
$480\leq t<780$& $\leq 4$& [1,2,3,4]       \\ \hline
$780\leq t<1080$& $\geq 7$&[12,13,14,15,16,17]         \\ \hline
$1080\leq t<1200$&    $>4$ and $\leq 7$& [5,6,7,8,9,10,11]          \\ \hline
$1200\leq t\leq 1440$& $\leq 4$&[1,2,3,4]       \\ \hline
\end{tabular}
\end{table}
The piecewise constant intensity function of the NHPP is constructed by randomly choosing a value for the arrival rate from the arrays given in the third column of Table \ref{tab2} and assuming it to be constant for short intervals of 30 minutes. The constant rate intensity function of the HPP is chosen to be $\lambda^{+}=\max {\lambda(t)}\forall t\in [0,T]$. The arrival times of the NHPP with a piecewise constant intensity function given in Table \ref{tab2} are obtained by applying the thinning algorithm to the considered HPP with the intensity function $\lambda^{+}$. Fig \ref{fig:3}  shows the arrival rate of the NHPP that models the arrivals of EVs queuing up at a charging station to receive charging service.  
\subsection{EV Consumption Profile}
The total EV demand is calculated based on the expected number of arrivals in a fixed interval ([0,24] hours) and the real world parameters such as daily driven miles, EV battery state of charge (SoC), EV battery capacity, consumption in kWh/100 miles, inverter capacity, charging rates and charging efficiency. As a first approximation, we consider a commercial building with EV charging facility where EVs queue to receive charging service. We also consider homogeneous EVs and use the battery parameters of Tesla Model 3 which is the most popular EV in North America \cite{Driver2021BestEVs}. \par
The daily driven miles can be modeled by a log-normal distribution with mean of 3.37 and a standard deviation of 0.5 \cite{Affonso2018ProbabilisticAging}. The battery state of charge (SoC) can be estimated from the daily driven miles (d), electricity consumption in kWh/100 miles ($E_{cons}$) and battery capacity ($C_{b}$) in kWh. 
\begin{equation}
    SoC_{arrival}=1-\frac{E_{cons}d}{C_b}\times 100
\end{equation}
The energy required to charge the battery to the desired SoC which in this case is $100\%$ is given by
\begin{equation}
    E_{req}=\frac{SoC_{final}-SoC_{arrival}}{\eta\times 100}C_b \label{eq:2.6}
\end{equation}

\begin{figure}[H]
    \centering
    \hbox{\hspace{-2.0em}}\includegraphics[width=3.5in,height=2.0in]{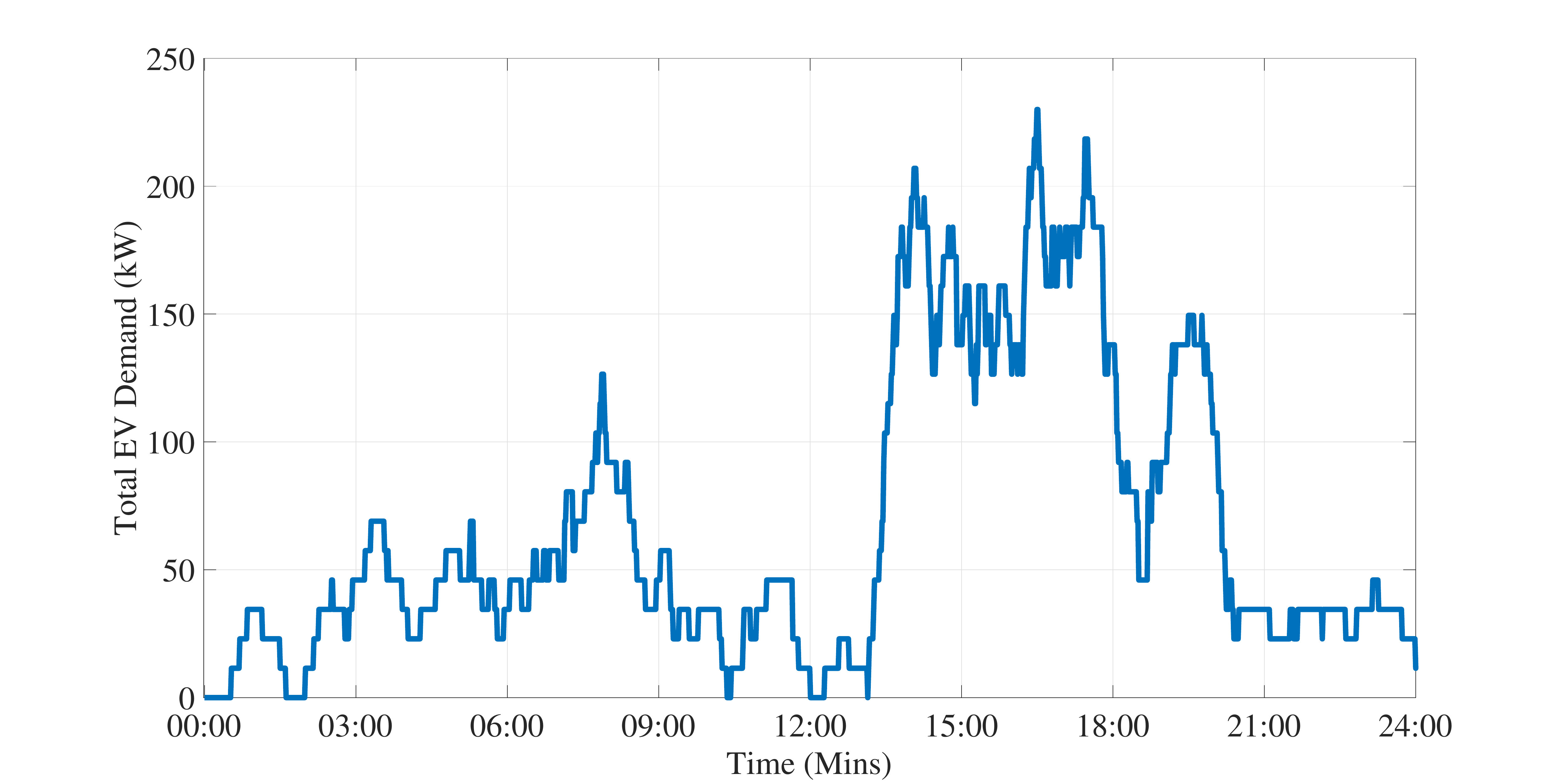}
    \vspace{-2.0em}\caption{EV Aggregate Demand}
    \label{fig:4}
\end{figure}

$\eta$ in (\ref{eq:2.6}) is the charging efficiency assumed to be $95\%$. The total charging time is given by
\begin{equation}
    T_{ch}=E_{req}/P \label{eq:2.7}
\end{equation}
$P$ in (\ref{eq:2.7}) is the charging rate. In this case we assume 3-$\phi$ L-2 charging. Table \ref{tab3} gives the battery related parameters of Tesla Model 3 considered in this work.
\begin{table}[H]
\centering
\caption{EV Battery Characteristics}
\label{tab3}
\vspace{-1.0em}
\begin{tabular}{|c|c|}
\hline
Parameter & Value          \\ \hline
$C_{b}$(kWhRated)& 75 kWh  \\ \hline  
P(kWRated)& 11.5 kW ; 48 Amps (3-$\phi$)       \\ \hline
$\eta $ & 95\%       \\ \hline
\% Charging& 100\% of kWRated        \\ \hline
$E_{cons}$&   27 kWh/100 miles          \\ \hline
\end{tabular}
\end{table}
The total EV demand is calculated based on the EV arrival rate as shown in Fig \ref{fig:3}, the expected number of events of the EV counting process $N_{EV}(t)$ in the interval $(0,t)$, the energy required to charge the EVs (\ref{eq:2.6}) and the total charging time given by (\ref{eq:2.7}). Based on the arrival rate shown in Fig \ref{fig:3}, the total EV electric demand over a period of one day is shown in Fig \ref{fig:4}

The EV demand profile in Fig \ref{fig:4} is characterized by sharp peaks due to the stochastic nature of the EV connection to the grid. This has the effect of increasing the peaky behavior of the aggregate load. The impact of EV charging such as shown in Fig \ref{fig:4} is discussed next.
\section{Aggregate Load Behavior with EV Charging}
The aggregate load of a power system has a strong temporal dependence. It has been shown that the power system demand is characterized by a lot of variability \cite{Singh2010StatisticalModel} with different load types following dissimilar patterns of use and as such cannot be statistically modeled by a single density function.  Since the connection of the EVs to the grid for charging purposes is inherently a stochastic process, the total electric demand is also characterized by an increased degree of randomness or uncertainty. Depending on the nature of the intensity function, the total EV demand can assume many scenarios with each scenario a realization of the stochastic counting process that models the EV arrivals. Also, the increased overall with intermittent sharp peaks can overload power equipment such as power transformers and transmission lines thereby increasing the probability of equipment failure due to increased thermal loading. In this work , we assume that the power system has enough capacity to support the EV charging scenario that is considered and all the physical power system constraints are satisfied.\par
\begin{figure}[H]
    \centering
    \hbox{\hspace{-2.0em}}\includegraphics[width=3.5in,height=2.0in]{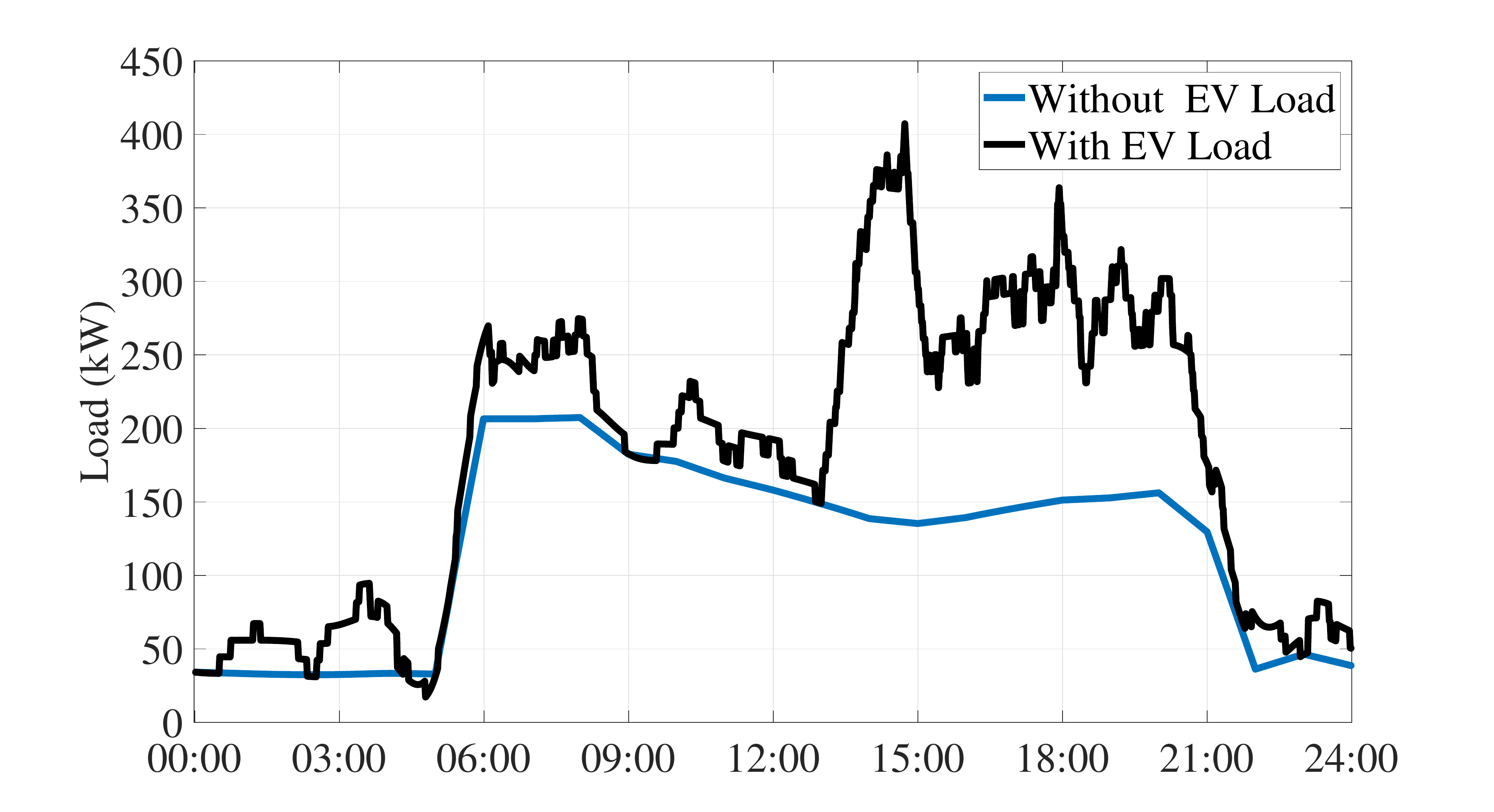}
    \vspace{-2.0em}\caption{Aggregate Load of Commercial Building}
    \label{fig:5}
\end{figure}
\begin{figure}[H]
    \centering
    \hbox{\hspace{-2.0em}}\includegraphics[width=3.5in,height=2.0in]{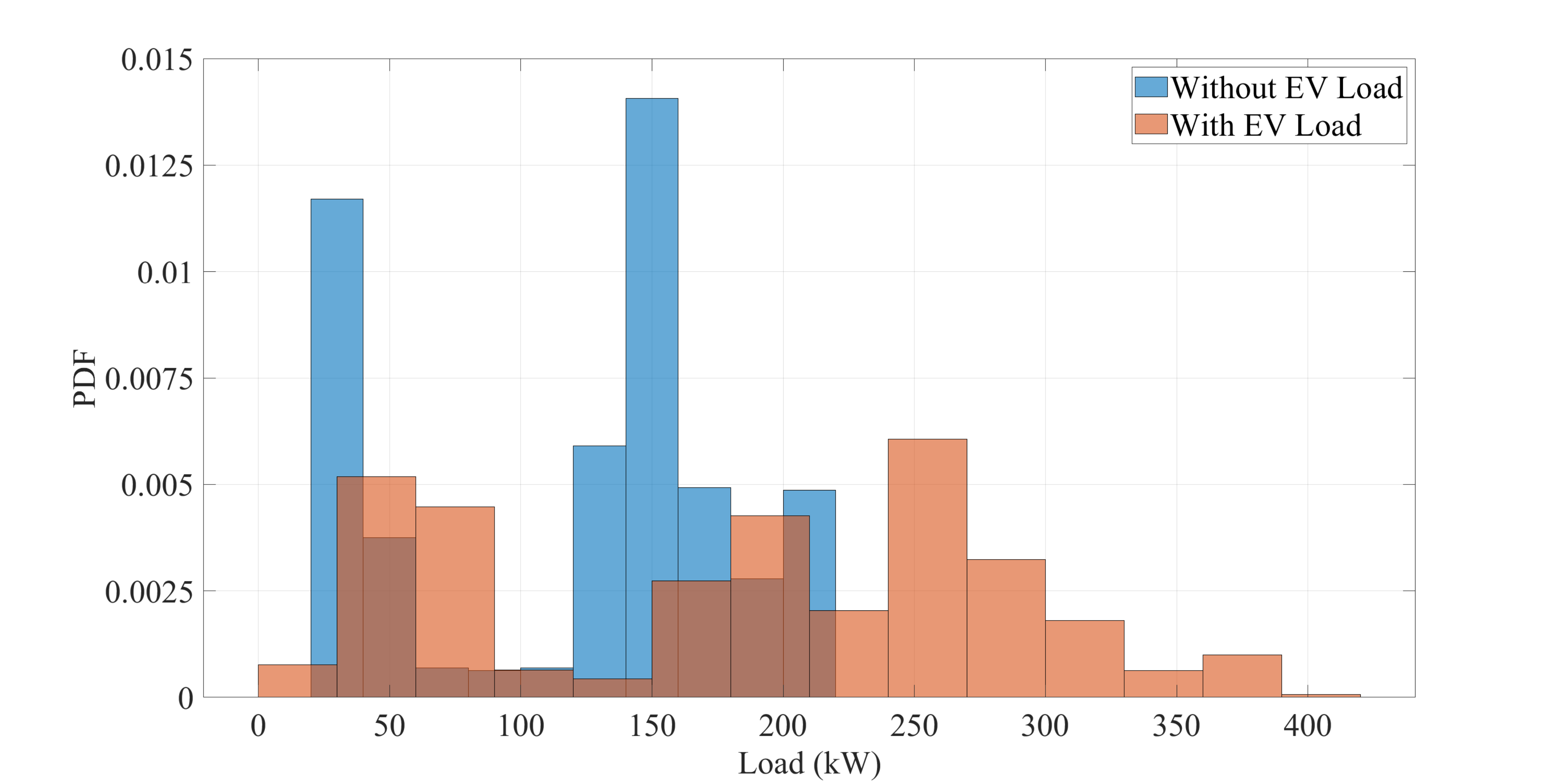}
    \vspace{-2.0em}\caption{PDF of Aggregate Load }
    \label{fig:6}
\end{figure}

The aggregate load of the commercial building with the EV charging facility is shown in Fig \ref{fig:5} . Two possible realizations of the total load as a function of time are shown. The blue curve is the load profile of the commercial building with zero EV charging. The black curve is the load profile of the same building considering EV charging. It can be seen that with EV charging, the overall demand has a lot of variability characterized by sharp rising and falling peaks. The peak demand without EV charging is 200 kW approximately. With EV charging the peak demand reaches 400 kW. Fig \ref{fig:6} shows the histogram of aggregate load for the two scenarios. It can be seen that EV charging introduces more peaks in the empirical density function of the aggregate load.  

\vspace{-2.5em}

\section{Generalized Gaussian Mixture Model}
One key limitation of a Gaussian distribution  is the lack of flexibility in modeling the statistical behavior of the loads that have a pronounced \say{peaky} characteristic. In that regard, it may be helpful to use a more general parametric model with some additional parameters that can adequately capture the impact of EV charging rates. One such model that offers more flexibility in modeling a large variety of statistical behaviors is a generalized Gaussian distribution . The generalized Gaussian distribution has one additional  parameter  referred to as the shape parameter. The value of this parameter controls the shape of the distribution with larger values resulting in  smaller tails and vice versa. For different shape parameter values, the generalized Gaussian can approximate a large class of other probability distributions. The generalized Gaussian we adopt in this paper has the density function of the following form
\begin{equation}
    f_{Y}(y;\mu,s,p)=\frac{\beta}{2s\Gamma\left(\frac{1}{\beta}\right)}\exp{\left[-\frac{|y-\mu|^{\beta}}{s^{\beta}}\right]} \label{eq:4.1}
\end{equation}
In \ref{eq:4.1}, $\beta \in \mathbb{R^+}$ is the shape parameter,$s\in \mathbb{R^+}$ is the scale parameter, $\mu\in \mathbb{R}$ is the location parameter and $\Gamma(.)$ is the gamma function. It is worth noting that $p=1$ gives a Laplace distribution and $p=2$ gives a Gaussian distribution. A mixture model based on the density function  (\ref{eq:4.1}) is a weighted convex combination of a finite number of generalized Gaussian distributions. Such a mixture model will be parameterized  by the weight, location, shape and scale parameter of each component density. In general, a mixture model that is a sum of a finite number of density functions has the following form
\begin{equation}
    f_{Y}(y_i;\Psi)=\sum_{j=1}^{M}\pi_j p_{j,Y}(y_i,C=j,\Theta_j) \label{eq:4.2}
\end{equation}
In the context of load modeling, $Y=y_i$ is the measured load data,  $M$ is the number of mixture components and $\pi_j=p(y_i\in C_{j})$ is the weight of the $j^{th}$ component density. The weights assigned to the component densities are subject to following constraints.
\begin{equation}
    \begin{aligned}
     &\pi_j>0 \;  \forall \: j \in \{1,2,...,M\}\\
    &\sum_{j=1}^{M} \pi_j=1  
    \end{aligned}
\end{equation}
Each component density function  $p_{j,Y}$ is characterized by the parameter vector $\Theta_j$ and $\Psi=\{\pi_j,\Theta_j ;j=[1,2,...,M]\}$ is the overall parameter vector. If $\{y_i\}_{i=1}^{N}$ is the measured data vector, then the likelihood function of the model in (\ref{eq:4.2}) has the form
\begin{equation}
    \mathbb{L}(\Psi|y_i)=\prod_{i=1}^{N}\sum_{j=1}^{M}\pi_j p_{j,Y}(y_i,C=j,\Theta_j) \label{eq:4.3}
\end{equation}
If we take the logarithm on both the sides of (\ref{eq:4.3}) we will get the log-likelihood function of the mixture model. 
\begin{equation}
    l(\Psi|y_i)=\sum_{i=1}^{N}\log\sum_{j=1}^{M}\pi_j p_{j,Y}(y_i|C=j,\Theta_j) \label{eq:4.4}
\end{equation}
The parameters that characterize the mixture model (\ref{eq:4.2}) can be obtained by the maximizing the log-likelihood function (\ref{eq:4.4}). The overall parameter vector $\Psi$ is the solution of 
\begin{equation}
    \Psi^{*}=\argmax_{\Psi} \sum_{i=1}^{N}\log\sum_{j=1}^{M}\pi_j p_{j,Y}(y_i|C=j,\Theta_j) \label{eq:4.5}
\end{equation}

The cost function in (\ref{eq:4.5}) is ill-posed and a solution cannot be obtained by the usual method of maximum likelihood estimation. A solution to (\ref{eq:4.5}) can be obtained however by using the Expectation-Maximization algorithm (EM) \cite{Sammaknejad2019AIdentification}. The E-M algorithm finds a solution by reinterpreting the measured data vector $\bold Y$ as incomplete data and assume the existence of unobserved data $\bold Z=\{z_{ji}\}_{i=j=1}^{i=N,j=M}$ that carry information about which density "generated" each data item $Y=\{y_i\}_{i=1}^{N}$. That is we assume 
\begin{equation}
  z_{ji}=\begin{cases}
    1 , y_{i}\in C_{j}\label{eq:q} \\
    0; y_{i} \notin C_{j} 
  \end{cases}
\end{equation}

With the introduction of hidden variables the complete data log-likelihood function assumes the following form
\begin{equation}
    l(\Psi|y_i,z_{ji})=\sum_{i=1}^{N} \log z_{ji} \pi_j p_{j,Y}(y_{i}|C=j,\Theta_j) \label{eq:4.6}
\end{equation}
Since the vector of hidden variables $\bold Z$ is unobserved and hence a random vector, the log likelihood function in (\ref{eq:4.6}) is a random variable and the E-M algorithm finds the lower bound of this expectation in the first step.  This is called the E-step of the algorithm. The next step involves maximizing the lower bound and the process repeats until the convergence criterion is met. \par
The expected value of the log-likelihood function as defined in (\ref{eq:4.6})  given the observed data and the current estimate of the parameter vector $\Psi^{k}$is 
\begin{equation}
    Q\left(\Psi^{k+1},\Psi^{k}\right)=\mathbb{E}\left[l(\Psi|\bold Y,\bold Z)|\bold Y,\Psi^{k}\right] 
\end{equation}
For the Generalized Gaussian Mixture model (GGMM) it can be shown that the expectation of the complete data log-likelihood function is given by
\begin{multline}
    Q\left(\Psi^{k+1},\Psi^{k}\right)=\sum_{i=1}^{N}\sum_{j=1}^{M}\mathbb{E}\left[z_{ji}|\bold Y,\Psi^{k}\right] \log \pi_j + \sum_{i=1}^{N}\sum_{j=1}^{M}\\\left(\log \beta_{j}-\log 2-\log s_j-\log \Gamma\left(\frac{1}{\beta_j}\right)-s_{j}^{-\beta_{j}}|y_i-\mu_j|^{\beta_j}\right) \label{eq:4.7}
\end{multline}
A key challenge in  calculating the expectation as defined in (\ref{eq:4.7}) is estimating $\mathbb{E}\left [z_{ji}|\bold Y, \Psi^{k}\right]$. Since the introduced hidden variables $\bold Z=\{z_{ji}\}_{i=j=1}^{i=N,j=M}$ are assumed to be binary-valued as defined in (\ref{eq:q}),  we can write for the expectation 
\begin{equation}
    \mathbb{E}\left [z_{ji}|\bold Y, \Psi^{k}\right ]=0.p\left(z_{ji}=0|\bold Y,\Psi^{k}\right)+1.p\left(z_{ji}=1|\bold Y,\Psi^{k}\right)
\end{equation}
In other words, this expectation is the probability that component density $j$ generated measurement $i$. This probability and hence the expectation can be calculated using Bayes' rule.
\begin{equation}
    \mathbb{E}\left [z_{ji}|\bold Y, \Psi^{k}\right ]=p\left(z_{ji}|\bold Y,\Psi^{k}\right)=\frac{p\left(y_{i}|z_{ji},\Psi^{k}\right)\pi_{j}^{k}}{\sum_{j=1}^{M} \pi_{j}^{k}p_{j,Y}(y_{i}|\Psi^{k})} \label{eq:4.8}
\end{equation}
The application of (\ref{eq:4.5})-(\ref{eq:4.8}) yields the following update equations  for the model parameters $\Psi^{(k+1)}=\{\pi_j,\mu_j,s_j,\beta_j;j=[1,2,...,M]\}$ of a GGMM given the current estimate $\Psi^{k}$ and the measured load data $\bold Y$ 
\begin{equation}
    \pi_j^{(k+1)}=\frac{1}{N}\sum_{i=1}^{N} \mathbb{E} \left[z_{ji}|\bold Y,\Psi^{(k)}\right]
\end{equation}
\begin{equation}
    \sum_{i=1}^{N}\mathbb{E}\left[z_{ji}|\bold Y,\Psi^{(k)}\right]\beta _{j}^{(k)}|\mu_{j}^{(k+1)}-y_{i}|^{\beta _{j}^{(k)}}=0
\end{equation}
\begin{equation}
    s_{j}^{(k+1)}=\left[\frac{\sum_{i=1}^{N}\mathbb{E}\left[z_{ji}|\bold Y,\Psi^{(k)}\right]}{\sum_{i=1}^{N}\mathbb{E}\left[z_{ji}|\bold Y,\Psi^{(k)}\right]\beta _{j}^{(k)}|\mu_{j}^{(k+1)}-y_{i}|^{\beta_{j}^{(k)}}}\right]^{-\frac{1}{\beta_{j}^{(k)}}}
\end{equation}
\begin{equation}
    \sum_{i=1}^{N}\mathbb{E}\left[z_{ji}|\bold Y,\Psi^{(k)}\right] \kappa=0 \label{eq:16}
\end{equation}
In (\ref{eq:16}), $\kappa$ equals
\begin{multline}
    \kappa= \frac{1}{\beta_{j}^{(k+1)}}+\frac{\psi\left(1/\beta_{j}^{(k+1)}\right)}{\left(\beta _{j}^{(k+1)}\right)^{2}}- \left(\frac{|y_i-\mu_{j}^{(k)}|}{s_{j}^{(k)}}\right)^{\beta_{j}^{(k+1)}} \\
    \left(\log |y_i-\mu_{j}^{(k)}|-\log s_{j}^{(k)}\right) \label{eq:17}
\end{multline}
In (\ref{eq:17}) $\psi(.)$ is the digamma function defined as $\Gamma ^{'}(g)/\Gamma(g)$. The update equations for the location and the shape parameter are nonlinear and we use an iterative solver like the Newton-Raphson to obtain a numerical solution. 

\section{Results and Discussion}
\begin{figure}[H]
\centering
\includegraphics[width=3.5in,height=1.7in]{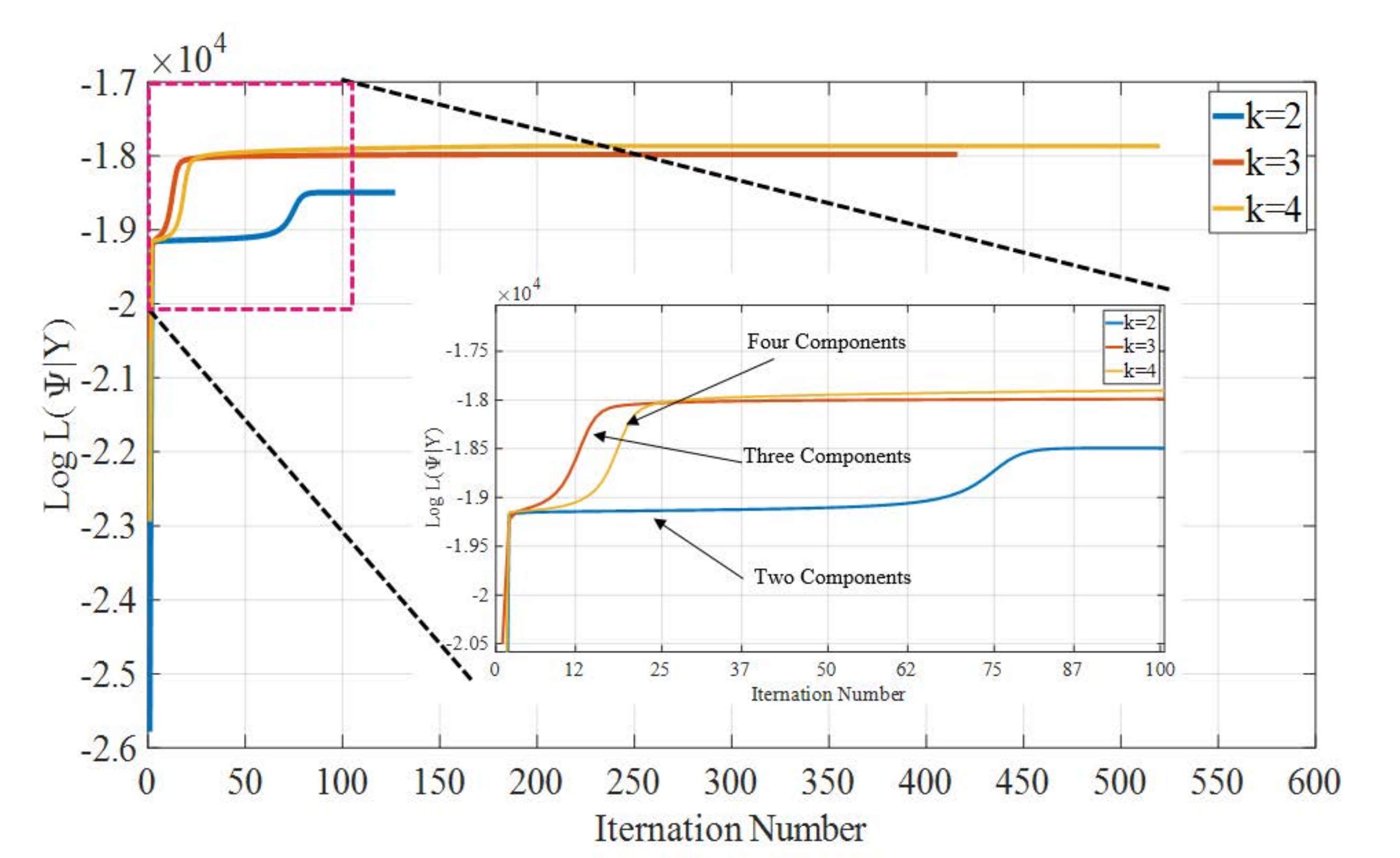}
\vspace{-1.5em}\caption{Changes in the Log-likelihood Function of GGMM}
\label{fig:7}

\includegraphics[width=3.5in,height=1.7in]{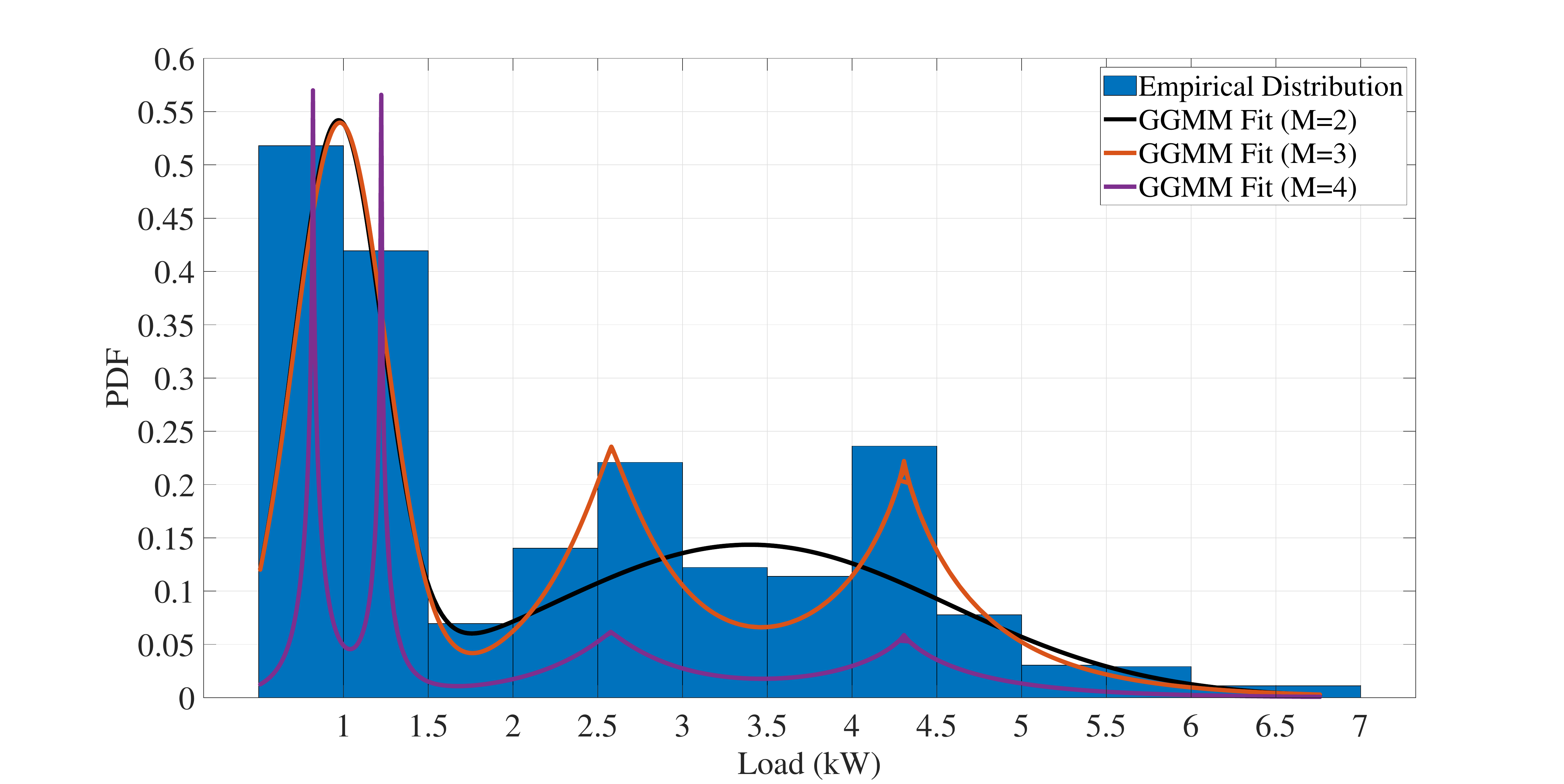}
\vspace{-1.5em}\caption{Load Distribution with GGMM Fit}
\label{fig:8}

\includegraphics[width=3.5in,height=1.7in]{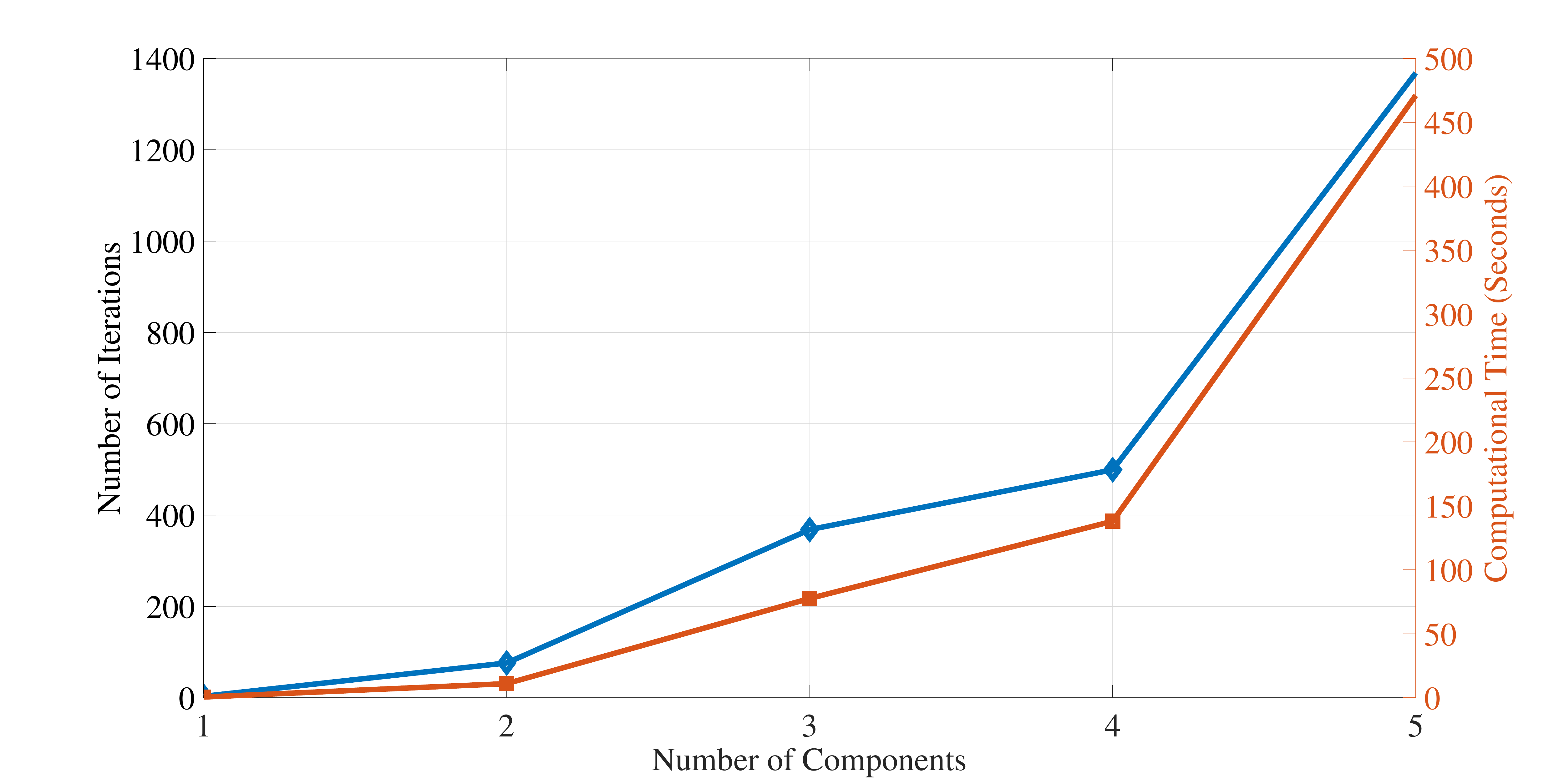}
\vspace{-1.5em}\caption{GGMM Performance Characteristics}
\label{fig:9}
\end{figure}

\begin{figure}
  \includegraphics[width=3.5in,height=1.7in]{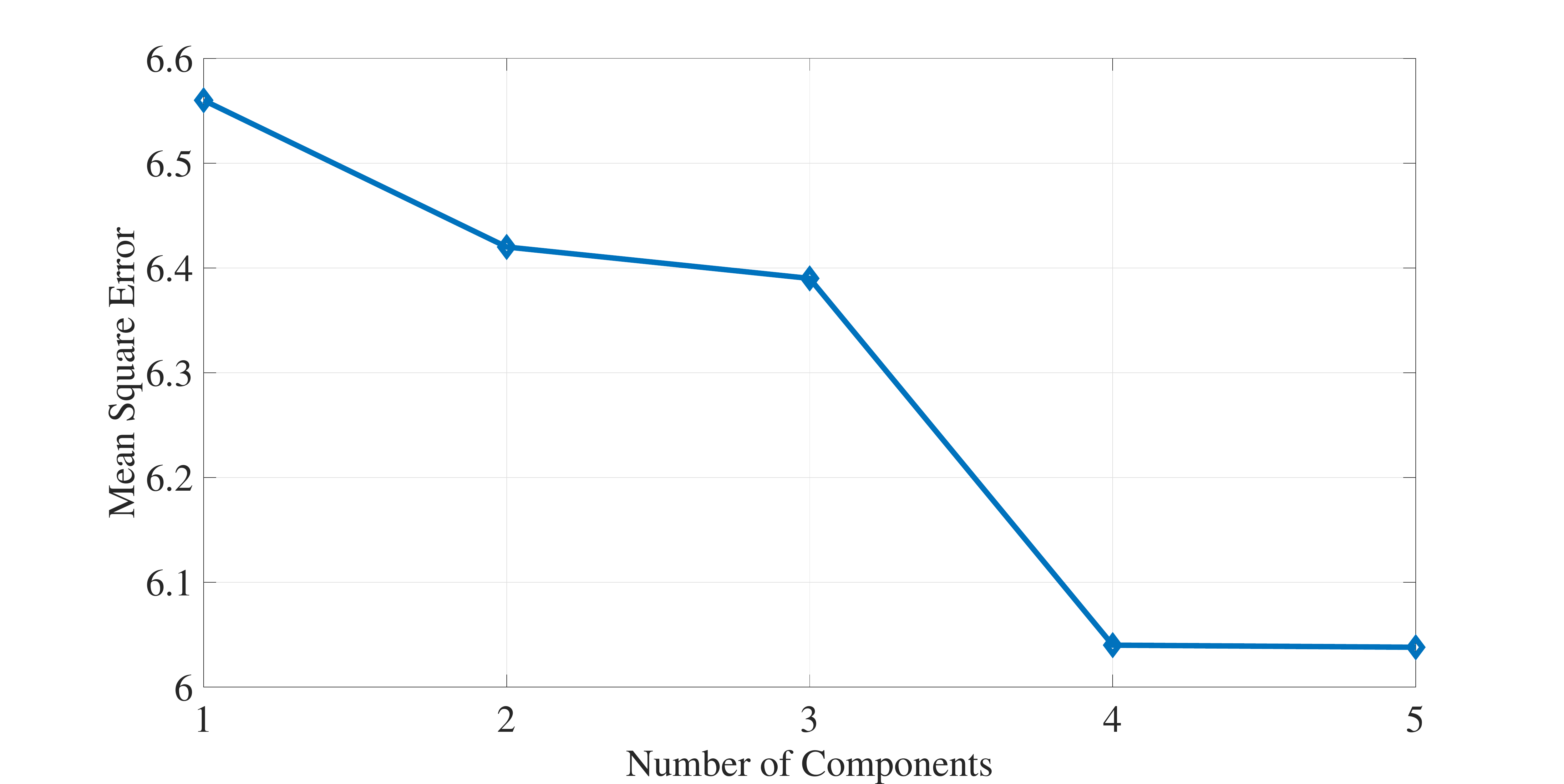}
\vspace{-1.5em}\caption{Mean Square Error of GGMM Fits}
\label{fig:10}  
 \end{figure}

In this section we present results of the univariate generalized Gaussian mixture model (GGMM) applied to the load distribution of a commercial building augmented with EV charging service. The base commercial load (total load without EV Charging) is obtained from the OpenEI website \cite{OfficeofEnergyEfficiencyRenewableEnergy2015CommercialDatasets}. The web-page was established by the Department of Energy (DoE) in 2009 and is a part of DoE mission to disseminate data in the public domain. We leave the multivariate formulation of the GGMM for the future work. The E-M algorithm used to obtain the parameters of the GGMM was coded in MATLAB \cite{MathWorks2020MATLABR2020b} and initialized using $\mathbf{K}-means$. The algorithm was run on an Intel Xeon processor with 32 GB of RAM. At each iteration of the algorithm the log-likelihood function is calculated and the difference between two consecutive values of the log-likelihood is compared with the tolerance. If the difference is less than tolerance, the algorithm is terminated. 
\begin{equation}
    \left|\frac{l^{k}-l^{k-1}}{l^{k-1}}\right|\leq \epsilon
\end{equation}
A threshold value of $\epsilon =1e-07$ was used to terminate the E-M algorithm. Fig \ref{fig:7} shows the log-likelihood function (\ref{eq:4.7}) of the GGMM plotted for different number of mixture components as a function of the iteration number. It is clear that the log-likelihood is nondecreasing at each iteration step and it can be seen from Fig \ref{fig:7}  that as the number of mixture components of the GGMM increases, the computational time to solve the model also increases. This can be inferred form Fig \ref{fig:7} since the log-likelihood takes more time to converge for a higher number of mixture components. The GGMM fit of the load distribution considering EV charging is shown in Fig \ref{fig:8}  . The GGMM model is able to capture the multi modal characteristics of the overall load distribution.  It is possible that a certain value of the load may be completely characterized by a single component of the GGMM in which that component may need to identified and isolated from the rest. In general, however a majority of the load values are best represented by some weighted combination of the mixture components. 
 From Fig \ref{fig:8} the GGMM density functions exhibit a pronounced peaky behavior as the number of component mixture densities increases. For $M=2$, the GGMM function has two distinct modes whereas for $M=4$ the resulting density function has four distinct modes. The fitting accuracy increases with the increase in $M$ but there is a price to pay in terms of the computational time the algorithm takes to converge.   The number of iterations required to estimate the parameters of the GGMM and the computational time in seconds are plotted as a function of number of components and the results are shown in Fig \ref{fig:9} .  On closer look the results shown in Fig \ref{fig:9} corroborate the changes in the log-likelihood function  as shown in Fig \ref{fig:7}   The  iteration count and the computational time share a nonlinear relationship with the number of mixture components.  If we increase the component densities for more accuracy, the algorithm  requires more iterations and hence more computational time to achieve convergence. There is no general rule for determining the optimal number of components in mixture models. For our application we use the mean square error metric to determine the optimal number of component densities. For the load histogram distribution shown in Fig \ref{fig:8}\par the mean square error as a function of number of components is plotted in Fig \ref{fig:10}. It is clear that there is no significant improvement in the fitting accuracy for $M>4$. For this reason we chose $M=4$ to generate the GGMM fit for the load distribution of Fig \ref{fig:8}. \par
 The GGMM fit can be used to generate new samples of load data with similar statistical properties. For example, to create a synthetic load profile, the GGMM can be fitted to the measured load data. From the fitted model, a random number generator function can be used to draw repeated samples. The new profile thus created will be statistically similar to the measured data. For the GGMM, a random number generator function can be obtained by taking the inverse of the CDF of the mixture model.
 \begin{equation}
    F_{Y}(y|\Psi)=\bigintsss_{-\infty}^{y}\sum_{j=1}^{M}\pi_j\frac{\beta_j}{2s_j\Gamma\left(\frac{1}{\beta_j}\right)}\exp{\left[-\frac{|t-\mu_j|^{\beta_j}}{s_j}\right]}dt \label{eq:6.1}
\end{equation}
A closed form solution of   $\hat{y}=F_{Y}^{-1}(y|\Psi)$ does not exit and hence numerical techniques such as Newton-Raphson must be used to generate random samples from the fitted model.

\section{ACKNOWLEDGMENT}
This publication was made possible by NPRP grant $\#$ 13S-0213-200357 from the Qatar National Research Fund (a member of Qatar Foundation). The statements made herein are solely the responsibility of the authors.

\section{Conclusion and Future Work}
This work presents a probability mixture model based on the generalized Gaussian distribution to develop a statistical model of the load considering L-2 EV charging. The EV charging profile is obtained by modeling the EV arrival process at a charging facility as a nonhomogeneous Poisson process. The NHPP model is used to estimate the arrival times of EVs. The charging time for each arriving EV is calculated from the daily driven miles and the state of charge of the EV at the time of arrival. The daily driven miles are assumed to be log-normally distributed, and the battery SoC at arrival is estimated considering the battery performance and parameters of Tesla Model 3. The EV charging profile thus obtained is added to the baseload profile of the commercial charging facility to obtain the overall demand pattern. The planning horizon considered is one day (24 hours).  \par
The statistical properties of the total electric demand with EV charging are modeled using a GGMM. The parameters of the GGMM are obtained using the E-M algorithm. The results shown in the paper demonstrate the applicability of a GGMM in representing load distribution with a pronounced peaky behavior. Since the proposed mixture model is parametric and hence \say{generative}. The model can generate synthetic load data with similar statistical properties as the measured data. Since the proposed model considers the EV charging load, the GGMM proposed has far-reaching applications such as probabilistic load flow with EV charging, distribution system state estimation (DSSE), where a large number of pseudo measurements are used to run the state estimation algorithms. The GGMM can model the non-gaussian distributed measurements, especially scenarios that involve a heavy penetration of EVs and distributed generation.\par 
Another important application of the proposed GGMM model is designing Monte Carlo simulations where the inputs are sampled from some underlying distributions. The GGMM model can be used to create different realizations of the electric demand curve and used as an input to run stochastic optimization algorithms. In the future, we aim to develop the multivariate GGMM and use the E-M algorithm to obtain the parameters of the multivariate GGMM. It is important to note that a multivariate formulation of the GGMM can be used to model the load correlation, which is not possible with a univariate formulation. The future work also involves extending the generalized formulation to other distributions such as beta prime and log-normal distributions and a performance comparison with the existing mixture models. 





%
\bibliographystyle{IEEEtran}
\bibliography{references1}

\begin{thebibliography}{10}
\providecommand{\url}[1]{#1}
\csname url@samestyle\endcsname
\providecommand{\newblock}{\relax}
\providecommand{\bibinfo}[2]{#2}
\providecommand{\BIBentrySTDinterwordspacing}{\spaceskip=0pt\relax}
\providecommand{\BIBentryALTinterwordstretchfactor}{4}
\providecommand{\BIBentryALTinterwordspacing}{\spaceskip=\fontdimen2\font plus
\BIBentryALTinterwordstretchfactor\fontdimen3\font minus
  \fontdimen4\font\relax}
\providecommand{\BIBforeignlanguage}[2]{{%
\expandafter\ifx\csname l@#1\endcsname\relax
\typeout{** WARNING: IEEEtran.bst: No hyphenation pattern has been}%
\typeout{** loaded for the language `#1'. Using the pattern for}%
\typeout{** the default language instead.}%
\else
\language=\csname l@#1\endcsname
\fi
#2}}
\providecommand{\BIBdecl}{\relax}
\BIBdecl

\bibitem{IEA2020GlobalIEA}
{IEA}, ``{Global EV Outlook 2020 – Analysis - IEA},'' 2020.

\bibitem{Room2021FactSheet}
\BIBentryALTinterwordspacing
W.~H.~B. Room, ``{Fact Sheet},'' 2021. [Online]. Available:
  \url{https://www.whitehouse.gov/briefing-room/statements-releases/2021/08/05}
\BIBentrySTDinterwordspacing

\bibitem{Elnozahy2014ANetworks}
M.~S. Elnozahy and M.~M. Salama, ``{A comprehensive study of the impacts of
  PHEVs on residential distribution networks},'' \emph{IEEE Transactions on
  Sustainable Energy}, 2014.

\bibitem{Clement-Nyns2010TheGrid}
K.~Clement-Nyns, E.~Haesen, and J.~Driesen, ``{The impact of Charging plug-in
  hybrid electric vehicles on a residential distribution grid},'' \emph{IEEE
  Transactions on Power Systems}, 2010.

\bibitem{Muratori2018ImpactDemand}
M.~Muratori, ``{Impact of uncoordinated plug-in electric vehicle charging on
  residential power demand},'' \emph{Nature Energy}, 2018.

\bibitem{Hilshey2013EstimatingAging}
A.~D. Hilshey, P.~D. Hines, P.~Rezaei, and J.~R. Dowds, ``{Estimating the
  impact of electric vehicle smart charging on distribution transformer
  aging},'' \emph{IEEE Transactions on Smart Grid}, 2013.

\bibitem{Alizadeh2014AVehicles}
M.~Alizadeh, A.~Scaglione, J.~Davies, and K.~S. Kurani, ``{A scalable
  stochastic model for the electricity demand of electric and plug-in hybrid
  vehicles},'' \emph{IEEE Transactions on Smart Grid}, 2014.

\bibitem{Hafez2018QueuingOperation}
O.~Hafez and K.~Bhattacharya, ``{Queuing analysis based PEV load modeling
  considering battery charging behavior and their impact on distribution system
  operation},'' \emph{IEEE Transactions on Smart Grid}, 2018.

\bibitem{Vlachogiannis2009ProbabilisticVehicles}
J.~G. Vlachogiannis, ``{Probabilistic constrained load flow considering
  integration of wind power generation and electric vehicles},'' \emph{IEEE
  Transactions on Power Systems}, 2009.

\bibitem{Garcia-Valle2009LetterStudies}
R.~Garcia-Valle and J.~G. Vlachogiannis, ``{Letter to the editor: Electric
  vehicle demand model for load flow studies},'' \emph{Electric Power
  Components and Systems}, 2009.

\bibitem{Bae2012SpatialDemand}
S.~Bae and A.~Kwasinski, ``{Spatial and temporal model of electric vehicle
  charging demand},'' \emph{IEEE Transactions on Smart Grid}, 2012.

\bibitem{Bricka2014NationalSurvey}
S.~Bricka and A.~Santos, ``{National Household Travel Survey},'' in
  \emph{Encyclopedia of Transportation: Social Science and Policy}, 2014.

\bibitem{Singh2010StatisticalModel}
R.~Singh, B.~C. Pal, and R.~A. Jabr, ``{Statistical representation of
  distribution system loads using Gaussian mixture model},'' \emph{IEEE
  Transactions on Power Systems}, 2010.

\bibitem{Valverde2012ProbabilisticModels}
G.~Valverde, A.~T. Saric, and V.~Terzija, ``{Probabilistic load flow with
  non-Gaussian correlated random variables using Gaussian mixture models},''
  \emph{IET Generation, Transmission and Distribution}, 2012.

\bibitem{Stephen2014EnhancedCustomers}
B.~Stephen, A.~J. Mutanen, S.~Galloway, G.~Burt, and P.~Jarventausta,
  ``{Enhanced load profiling for residential network customers},'' \emph{IEEE
  Transactions on Power Delivery}, 2014.

\bibitem{Quiros-Tortos2018StatisticalApplications}
J.~Quiros-Tortos, A.~Navarro-Espinosa, L.~F. Ochoa, and T.~Butler,
  ``{Statistical representation of EV charging: Real data analysis and
  applications},'' in \emph{20th Power Systems Computation Conference, PSCC
  2018}, 2018.

\bibitem{Cui2018StatisticalModel}
M.~Cui, C.~Feng, Z.~Wang, and J.~Zhang, ``{Statistical representation of wind
  power ramps using a generalized Gaussian mixture model},'' \emph{IEEE
  Transactions on Sustainable Energy}, 2018.

\bibitem{Sammaknejad2019AIdentification}
N.~Sammaknejad, Y.~Zhao, and B.~Huang, ``{A review of the Expectation
  Maximization algorithm in data-driven process identification},'' 2019.

\bibitem{Papoulis1967Processes}
A.~Papoulis and J.~G. Hoffman, ``{ Probability, Random Variables, and
  Stochastic Processes },'' \emph{Physics Today}, 1967.

\bibitem{Pasupathy2011GeneratingProcesses}
R.~Pasupathy, ``{Generating Nonhomogeneous Poisson Processes},'' in \emph{Wiley
  Encyclopedia of Operations Research and Management Science}, 2011.

\bibitem{Lewis1979SIMULATIONTHINNING.}
P.~A. Lewis and G.~S. Shedler, ``{SIMULATION OF NONHOMOGENEOUS POISSON
  PROCESSES BY THINNING.}'' \emph{Naval research logistics quarterly}, 1979.

\bibitem{Chen2016ThinningProcesses}
\BIBentryALTinterwordspacing
Y.~Chen, ``{Thinning Algorithms for Simulating Point Processes},'' Tech. Rep.,
  2016. [Online]. Available:
  \url{https://www.math.fsu.edu/~ychen/research/Thinning algorithm.pdf}
\BIBentrySTDinterwordspacing

\bibitem{Driver2021BestEVs}
\BIBentryALTinterwordspacing
C.~a. Driver, ``{Best Selling EVs},'' 2021. [Online]. Available:
  \url{https://www.caranddriver.com/features/g36278968/best-selling-evs-of-2021/}
\BIBentrySTDinterwordspacing

\bibitem{Affonso2018ProbabilisticAging}
C.~M. Affonso and M.~Kezunovic, ``{Probabilistic assessment of electric vehicle
  charging demand impact on residential distribution transformer aging},'' in
  \emph{2018 International Conference on Probabilistic Methods Applied to Power
  Systems, PMAPS 2018 - Proceedings}, 2018.

\bibitem{OfficeofEnergyEfficiencyRenewableEnergy2015CommercialDatasets}
{Office of Energy Efficiency {\&} Renewable Energy}, ``{Commercial and
  Residential Hourly Load Profiles for all TMY3 Locations in the United States
  - Datasets - OpenEI Datasets},'' 2015.

\bibitem{MathWorks2020MATLABR2020b}
T.~MathWorks, ``{MATLAB (R2020b)},'' \emph{The MathWorks Inc.}, 2020.

\end{thebibliography}

\end{document}